\documentclass[runningheads]{llncs}
\usepackage[T1]{fontenc}
\usepackage{graphicx}
\usepackage{hyperref}
\usepackage{xcolor}
\hypersetup{
    colorlinks,
    linkcolor={black},
    citecolor={black},
    urlcolor={black}
}

\begin{document}
\title{Pied Piper: Meta Search for Music}
\author{Pulak Malhotra\inst{1}\orcidID{0000-0002-2492-7981} \and
Ashwin Rao\inst{1}\orcidID{0000-0002-1545-2611}}
\authorrunning{Malhotra, P. and Rao, A.}
\institute{International Institute of Information Technology, Hyderabad, India
\email{\{pulak.malhotra,ashwin.rao\}@students.iiit.ac.in}}
\maketitle              % typeset the header of the contribution
\begin{abstract}
Internet search engines have become an integral part of life, but for pop music, people still rely on textual search engines like Google. We propose Pied piper, a meta search engine for music. It can search for music lyrics, song metadata and song audio or a combination of any of these as the input query and efficiently return the relevant results.

\end{abstract}
\section{Introduction}
We propose Pied Piper, a new search engine solely for pop music. Pied Piper works on 3 axes: lyrics, metadata and audio. Traditional text based search engines can only identify songs from the title or from lyrics. This is not ideal. For example, it could be hard to specify if we want a given word to be in the title, genre, artist name, or lyrics. Another case could be that if we have an audio snippet of an unknown song which we know was released before 2000, we cannot use both the audio snippet and the fact that it was released before 2000 in the same search. Pied piper solves this by using the concept of meta search along the 3 aforementioned axes to deliver the best search results in an efficient manner. We detail Pied Piper's methodology to search on lyrics, audio and metadata and how to combine rankings on each of these axes to obtain a final ranking of songs based on the user's input query. We also address error handling in audio matching and explain Pied Piper's software architecture and efficient parallel pipeline for music search. Finally, we suggest ways to evaluate performance and propose additional features that can be integrated into Pied Piper's existing design.

\section{Related Work}
A survey on audio fingerprinting \cite{cano} characterized all techniques of audio fingerprinting using a general framework. Improvements to resilience made systems such as Shazam \footnote{https://www.shazam.com/} \cite{wang}, Philips \cite{Haitsma2002AHR}, and Microsoft \cite{burges} popular, especially for the music information retrieval task. Google has released popular audio recognition features in recent years. Now Playing \cite{google-ai} is a Shazam-like functionality on Pixel phones that recognizes songs from audio using an on-phone deep neural network. Sound Search \cite{google-ai} expanded this functionality to support over 100 million songs. The new hum-to-search feature on Google Assistant allows a user to hum the tune of a song and matches this against Google's database of songs using audio fingerprints. \cite{samaf} proposed SAMAF, which uses a sequence-to-sequence autoencoder model consisting of long-short term memory (LSTM) \cite{lstm} layers to generate audio fingerprints. SAMAF was shown to perform better than other audio fingerprinting techniques in the same paper.

\section{Problem Formulation}

\subsection{Lyrics and Metadata Search}
Lyrics search refers to the retrieval of all songs which contain a given string in their lyrics, ranked by relevance. Similarly, we define metadata search as the retrieval of songs which have specific fields in their metadata, matching with fields in the given query. Lyrics and metadata search about music can be tackled as a text-based domain-specific search problem. Metadata parameters can include information like music name, album name, artist name and release year.

\subsection{Audio Search}
Audio search refers to the retrieval of all songs which contain the exact or perceptually similar audio to the given audio query, ranked by relevance. Audio search is an entirely different problem compared to text search due to audio being high dimensional. Perceptually similar audio samples could have substantial variance in their data streams. As a result, the comparison of one audio sample against all other audio samples is computationally very expensive and neither efficient nor effective. An \textit{audio fingerprint} is a content-based compact signature that summarizes essential information about the audio sample. Instead of using the entire audio file for comparison, we use the audio fingerprint instead.

Traditionally, a meta search engine is defined as a search engine which collects results from other search engines (such as Google and Bing) and then makes a new ranking. We use this term in a slightly different way, as the independent search engines are internal parts of Pied Piper (we will discuss this in detail in Sec. \ref{sec:system_architecture}). We run all searches for different fields in the input independently. This allows us to run them concurrently and deliver results as quickly as possible. An alternative we considered was using a transformer \cite{trans} and embedding a multi-dimensional query (for example, a query with song snippet, song name and artist name) in a contextualized feature space, but this would turn out to be too slow as audio information is itself is high dimensional and difficult to encode.

\section{System Architecture}
\label{sec:system_architecture}

\begin{figure}[h]
\centering
\includegraphics[width=\columnwidth]{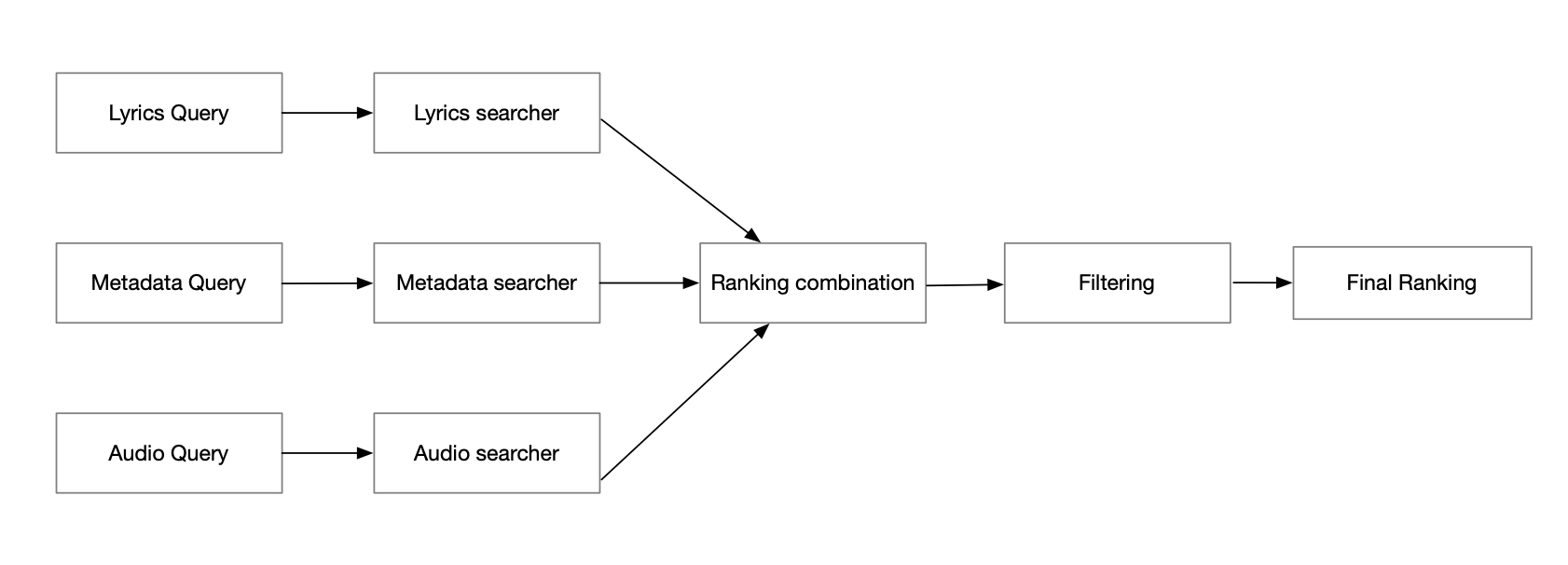}
\caption{System Architecture}
\label{fig:fig_system_architecture}
\end{figure}

Pied piper has a frontend which accepts user query. The query can be divided along the 3 axes as mentioned earlier. Now, each sub-query is parsed and processed to generate a ranking and score for the document. All this information is passed into a merging step to generate the combined ranking, which, after filtering, gives us the final ranking. The system architecture is shown in Fig. \ref{fig:fig_system_architecture}.

\subsection{Frontend}
The frontend of Pied Paper has an easy-to-use user interface. There is one input bar for lyrics, one button to record audio or upload an audio file, and a few more inputs for different kinds of metadata that the user knows. It is not compulsory for the user to fill up all the fields. The more fields the user fills, the more information is available to Pied Piper, leading to better results.

\begin{figure}[h]
\centering
\includegraphics[width=0.9\columnwidth]{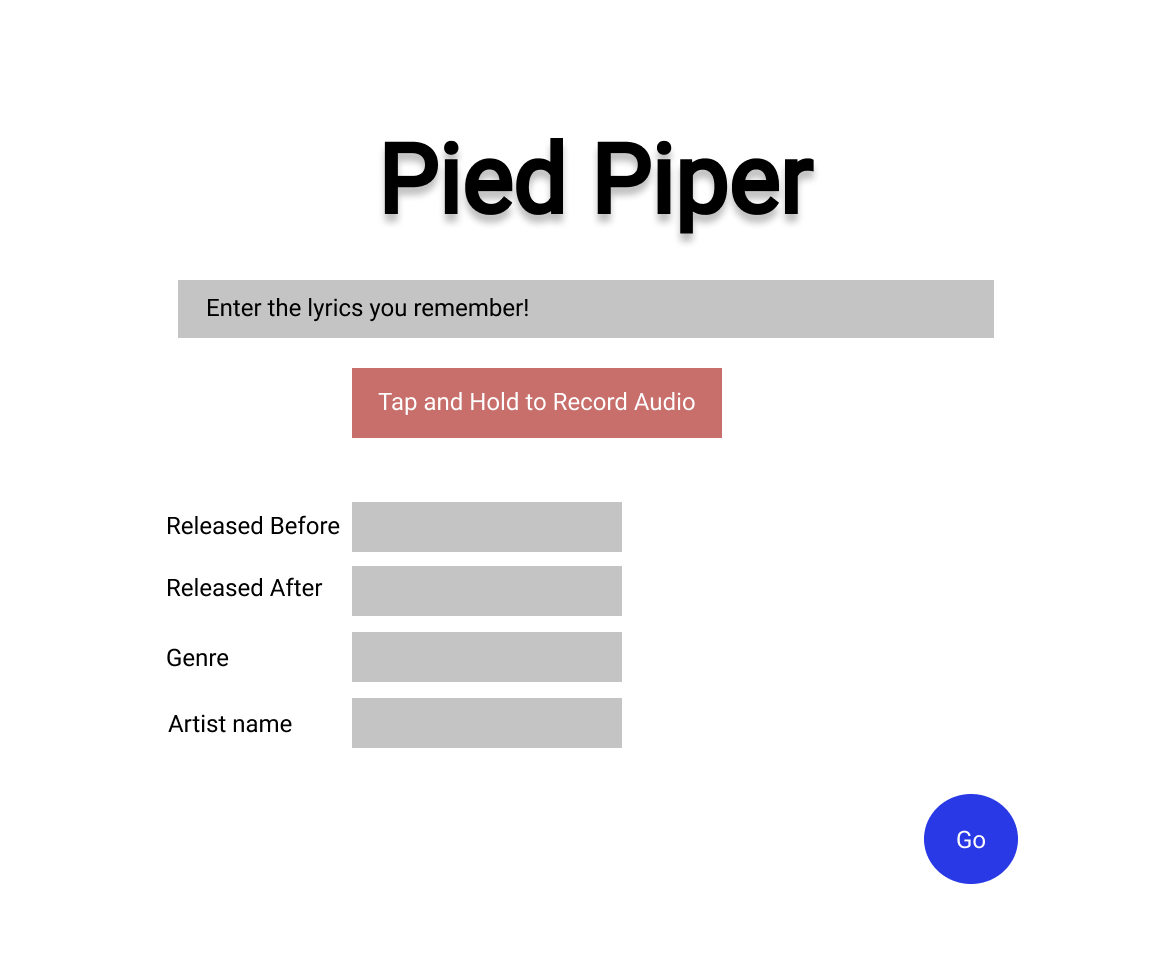}
\caption{A Sample UI for the frontend of Pied Piper. User can enter lyrics, record audio, search by artist name and can get songs released before or after a given date.}
\end{figure}

\subsection{Backend}

\subsubsection{Database}
The backend of the system is assumed to have a database of all songs along with all metadata of every song. This information can be easily scraped from a website like Youtube \footnote{\url{https://youtube.com}} or Soundcloud \footnote{\url{https://soundcloud.com/}}. MusicBrainz \footnote{\url{https://musicbrainz.org/}} is an open source music encyclopedia which can be used to collect metadata about songs.

\subsubsection{Lyrics Search}
For lyrics search, we use a text search engine using an inverted index (posting list) to allow for fast searches. The index contains the tokens and the $N$ gram of lyrics as keys. Stop-words in lyrics are removed since they play no role in song identification. Each posting list will contain the song ID of the songs that contain those tokens. The value of $n$ can be tuned according to the space requirements. If space is a constraint, the value of $n$ can be reduced, at the expense of a slight decrease in accuracy. This is a trade-off between memory and processing.

\begin{figure}[h]
\centering
\includegraphics[width=0.7\columnwidth]{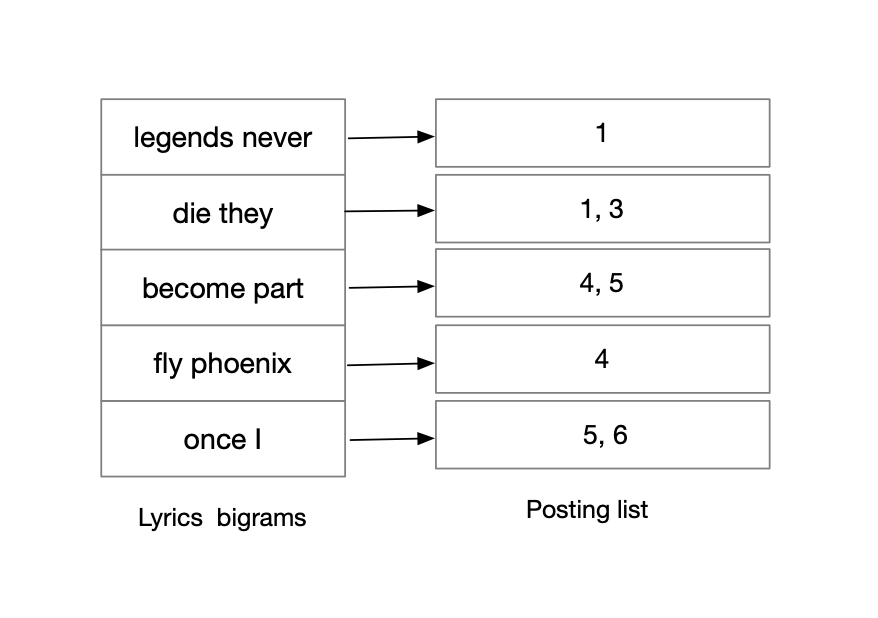}
\caption{Sample inverted index for lyrics}
\end{figure}

A term-frequency inverse document frequency (tf-idf) ranking system would work well in the case of lyrics. As the catchy parts of the songs are often repeated many times in a given song (like the chorus), the term frequency would be higher and would boost the ranking of relevant songs when they are searched for using these repeated portions of lyrics.

\subsubsection{Metadata Search}
\label{sec:metadata_search}
Metadata search for textual features like track title is implemented in a similar fashion to lyrics search, but with minor modifications. For example, in case of title search, stop words are not removed since titles are usually short and stop words could have important meaning. For features like the release date, metadata search can act as a filter on the final results. If the input for date is before 2008, all the songs released after 2008 in the search results can simply be removed. For each type of metadata, a different ranking and document score is returned. Hence, search on each metadata type is treated as an independent process. The rankings for each metadata type will be merged in the merging step.

\subsubsection{Audio Search}
\begin{figure}[h]
\centering
\includegraphics[width=0.7\columnwidth]{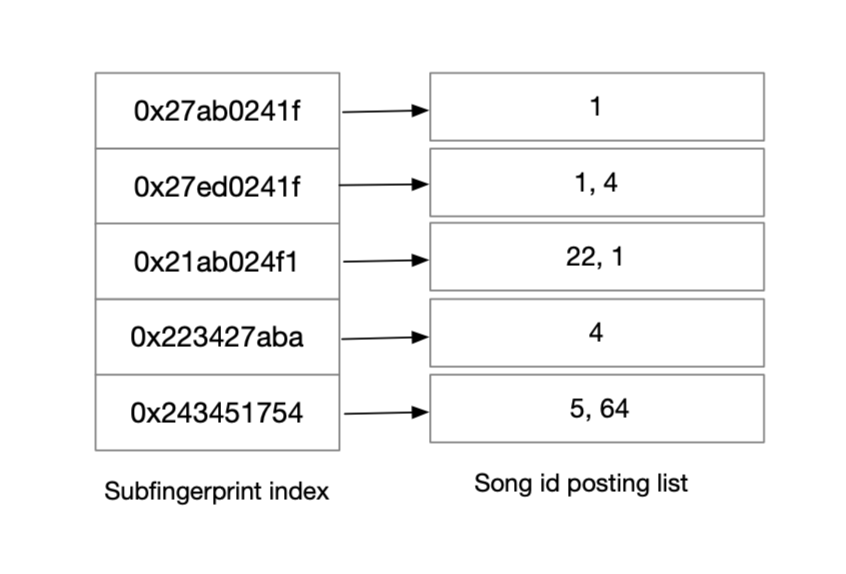}
\caption{Sample inverted index for audio fingerprints}
\label{fig:fingerprint_index}
\end{figure}

We propose the use of SAMAF \cite{samaf} to generate a set of 32 bit sub-fingerprints for a song (after breaking the song into many small-sized intervals). We store all these sub-fingerprints in the database and need an effective way to index them. \cite{cha} proposed an effective way to index audio fingerprints. We adopt a similar search methodology and use an inverted index. Each sub-fingerprint is mapped to a list of songs that contain the given sub-fingerprint. A sample inverted list is shown in Fig. \ref{fig:fingerprint_index}.

\begin{figure}[h]
\centering
\includegraphics[width=\columnwidth]{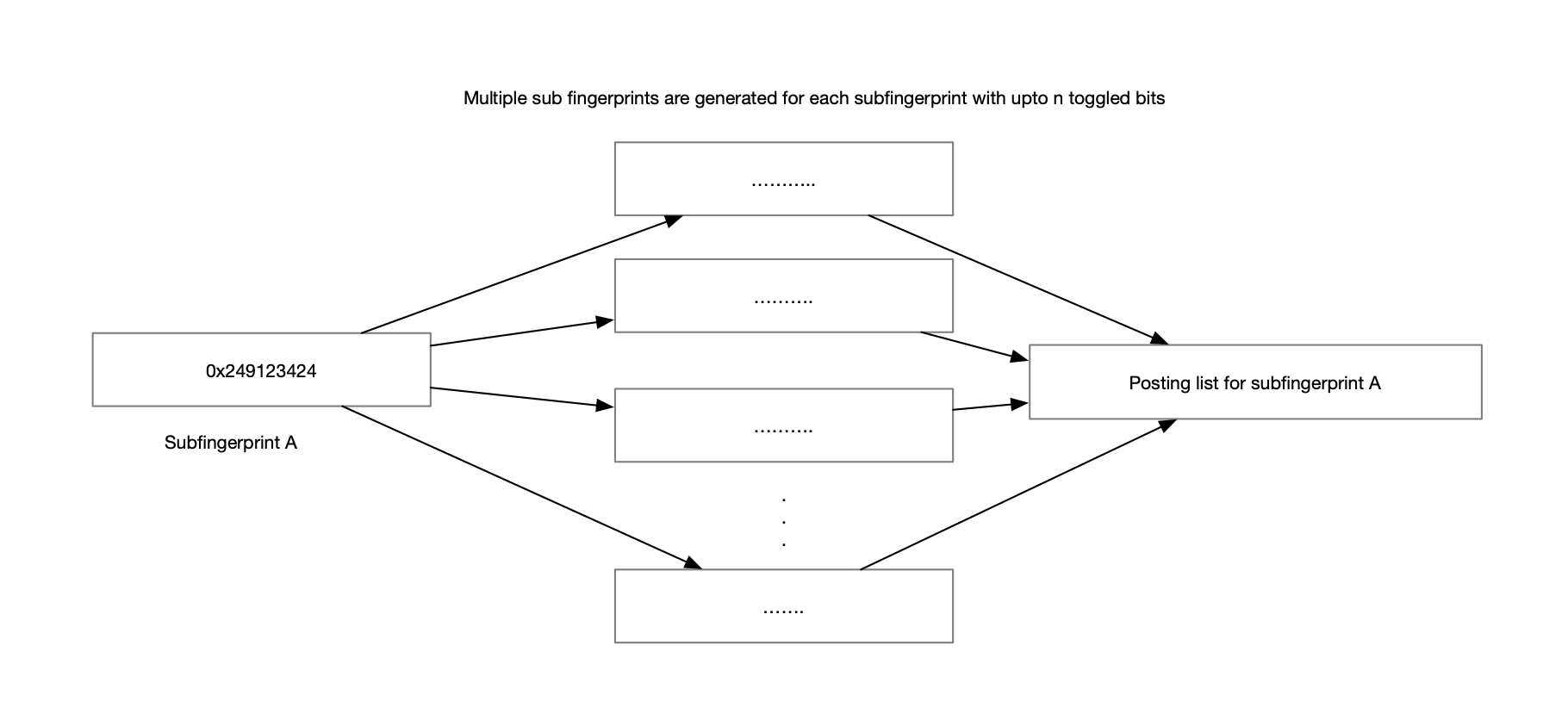}
\caption{The fingerprints generated with up to $n$ toggled bits point to the same posting list as that for fingerprint A.}
\label{fig:fingerprint_fig_2}
\end{figure}

For each sub-fingerprint, we also generate $\sum_{i=1..n} {32 \choose i}$ sub-fingerprints to account for up to $n$ toggled bits. Generating these fingerprints with toggled bits avoids situations when matching fails due to erroneous bits in the query fingerprint. Fig. \ref{fig:fingerprint_fig_2} shows how the inverted index looks after adding the generated sub-fingerprints. Adding the toggled bit increases the index size; however, we can achieve $O(1)$ lookup time with B+ Trees.

\begin{figure}[h]
\centering
\includegraphics[width=\columnwidth]{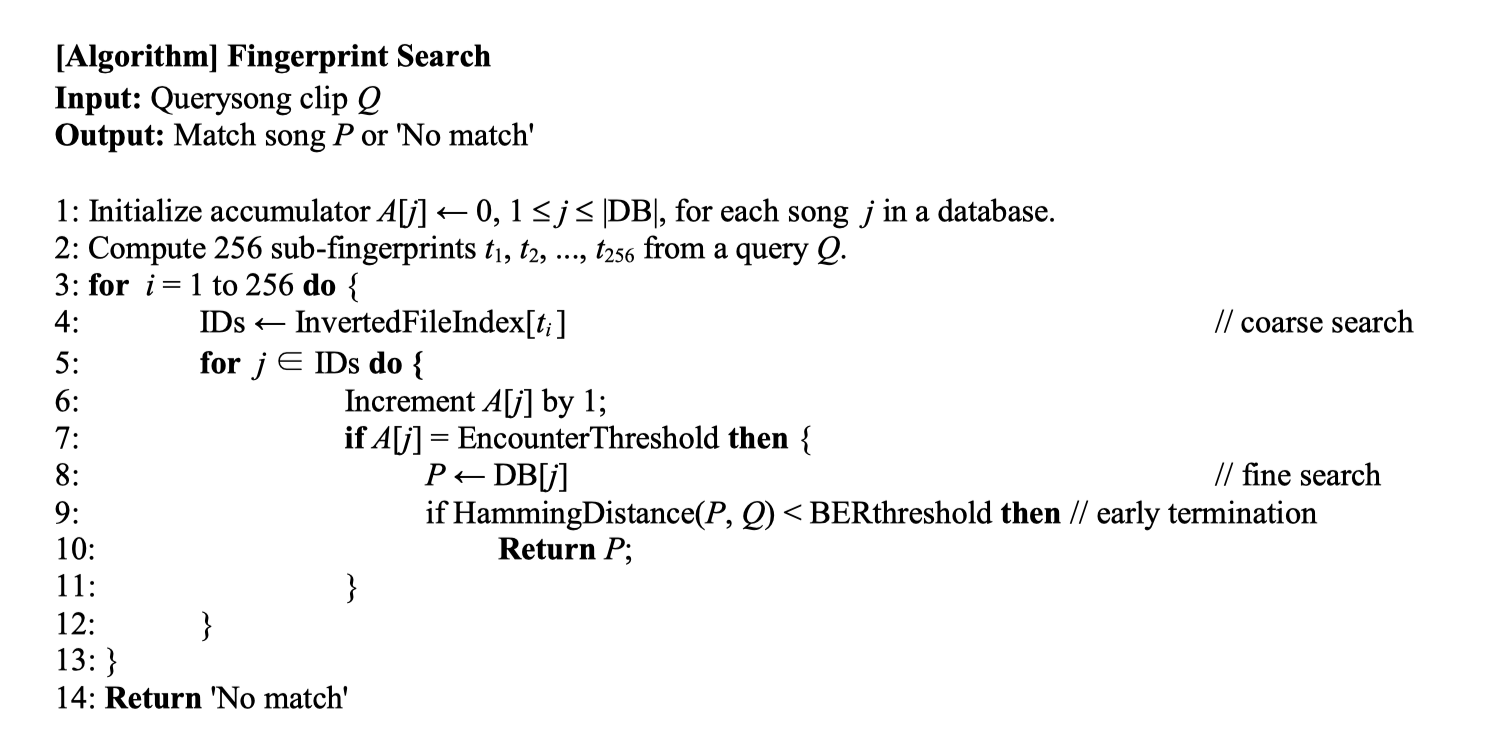}
\caption{Search algorithm for audio fingerprints \cite{cha}}
\label{fig:search-algo}
\end{figure}

The search with this index is done in a two-fold manner. First, there is a \textit{coarse} search in which the all songs associated with the sub-fingerprints of the query are retrieved. Then, for the songs which have large number of sub-fingerprints matching, a \textit{fine} search is performed. In \textit{fine} search, the full fingerprint of the song is retrieved (this is a heavy operation) and the similarity between the retrieved and query fingerprint is calculated using Hamming distance between them. If the similarity is above a particular threshold, the song is reported as a result. The algorithm to search for audio fingerprints is shown in Fig. \ref{fig:search-algo}.

\subsection{Merging and creating the final ranking}
Once all 3 types of searches are complete, each of them returns a set of results along with a normalized score for each song in the result. We can run the aforementioned searches in parallel as they are independent of each other. This will boost the speed of the search engine significantly. 
If the user only enters one kind of input of query, then the result of that particular search can be shown as the final result.
If the user has specified more than one type of field (such as both lyrics and audio) then we need to merge the results. We propose the use of weighted sum for this purpose. Suppose the score for a song $r$ in the $i$th field is given by $r_i$ then,

\begin{equation}
    FinalScore(r) = \sum_{i=1}^{n}{c_i * r_i}
\end{equation}
where $\sum_{i}^{n}{c_i} = 1$. \\

The value of $n$ depends on the number of features the user's input has. For example, if the user gives an audio sample, artist name, track name and lyrics, then the value of $n$ will be $4$. The values of $c_i$ can act as hyperparameters which we can tune. In general, a feature like track name should have a higher weight than lyrics since we are less likely to find false positive matches in the song's title as compared to its lyrics.

The last step of creating the final ranking is to remove those results which do not fit a metadata filter (as discussed in Sec. \ref{sec:metadata_search}). After filtering, we sort the results by their final scores in descending order and return the results to the user on the frontend.

\section {Evaluation}
Each of the field searches can be individually evaluated by providing inputs only for that field. This can be used to tune the value of $N$ in the $N$-gram index that we make for textual searches. For audio search, the thresholds in both \textit{fine} and \textit{coarse} search can be treated as hyperparameters and tuned to get better results. We can add white noise in various amounts to song recordings to test what level of noise starts creating erroneous results. Further, since the SAMAF framework uses deep learning, we could try to generate adversarial examples using the weights of the model.

\section{Conclusion and Future Work}
In this paper, we propose a meta search engine for music which can effectively handle multi-dimensional input, be it lyrics, track metadata or track audio. There is always scope for improvement. The robustness of SAMAF to certain audio distortions such as pitch shifting and speed change is still weak, and other audio fingerprinting techniques which are more robust to these factors can be utilized. Oftentimes, people remember sentiments of the song. For example, whether the song was energetic or sad. It could be interesting to classify songs into different sentiments and themes using machine learning techniques and provide an option for such sentiment matching in the search engine as well. Another aspect to work on could be the personalization of the search engine based on users' song listening history.

% \clearpage

% \bibliography{references/refs}





% LNCS Homepage, \url{http://www.springer.com/lncs}. Last accessed 4
% \end{thebibliography}

\end{document}